\documentclass[prb,twocolumn,showpacs,preprintnumbers]{revtex4}

\usepackage{graphicx}
\usepackage{amsmath}
\usepackage{amssymb}

\begin{document}

\title{Decoherence of coherent transport in a disordered
one-dimensional wire: Phenomenological model}

\author{Martin Mo\v{s}ko}
\email{elekmoso@savba.sk}
\affiliation{Institute of Electrical Engineering, Slovak Academy of
Sciences, D\'{u}bravsk\'{a} cesta 9, 842 39 Bratislava, Slovakia}

\author{Pavel Vagner}
\affiliation{Institut f\"{u}r Schichten und Grenzfl\"{a}chen,
Forschungszentrum J\"{u}lich GmbH, 52425 J\"{u}lich, Germany}
\affiliation{Institute of Electrical Engineering, Slovak Academy of
Sciences, D\'{u}bravsk\'{a} cesta 9, 842 39 Bratislava, Slovakia}

\author{Peter Marko\v{s}}
\affiliation{Institute of Physics, Slovak Academy of Sciences,
D\'{u}bravsk\'{a} cesta 9, 842 28 Bratislava, Slovakia}

\author{Thomas Sch\"{a}pers}
\affiliation{Institut f\"{u}r Schichten und Grenzfl\"{a}chen,
Forschungszentrum J\"{u}lich GmbH, 52425 J\"{u}lich, Germany}

\date{\today}

\begin{abstract}
We model the effect of phase-breaking collisions on the coherent
electron transport in a disordered one-dimensional single-channel
wire. In our model the phase-breaking collisions break the wire
into segments, where each segment is an independent series
resistor with coherent electronic resistance and the segmentation
is a stochastic process with Poisson distribution of
phase-breaking scattering times. The wire resistance as a function
of the wire length $L$, coherence length $L_{\phi}$, and
localisation length $\xi$ is calculated and the transition from
coherent to incoherent transport is traced quantitatively. In the
coherent regime ($L\!<\!L_{\phi}$) the resistance fluctuates from
wire to wire with a characteristic log-normal distribution of
resistances, the typical resistance increases as $\exp(L/\xi)$,
and the mean resistance increases as $\exp(2L/\xi)$ (or faster if
disorder is strong). As $L$ exceeds $L_{\phi}$, decoherence
suppresses the resistance fluctuations and narrows the resistance
distribution. As a result, at $L\!\gg\!L_{\phi}$ the mean
resistance increases as $\beta L-c$ and the typical resistance as
$\beta L- c^\prime $, where $\beta$ is the wire resistivity, $c$
is a constant shift due to the decoherence near the source
electrode, and $c^\prime \gg c$ is the shift related to the
resistance self-averaging in a single wire. Numerical results are
given for a GaAs quantum wire. It is noted that coherent transport
in such wire can exhibit peculiar deviations from universal scaling owing
to strong backscattering by impurities.
\end{abstract}


\maketitle

\section{Introduction}

Electron gas confined in a GaAs quantum wire is a realistic
one-dimensional (1D) electron gas system that almost ideally
manifests (or is expected to manifest) a variety of fundamental 1D
transport effects. \cite{Weisbuch-91}

If the wire is much shorter than the electron mean free path,
electron transport is ballistic. Then each energy subband occupied
by the 1D electrons contributes to the wire conductance by an
amount $2e^2/h$, i.e., the conductance is universally quantized
and wire-length-independent. \cite{PointContact,Tarucha-95} This
effect is explained by the Landauer conductance formula
\cite{Landauer-70,Landauer-90,Datta-95} if one considers just
ballistic transport of non-interacting 1D electrons. Incorporation
of many-body effects gives the same result. \cite{Maslov}

Disorder affects the 1D transport in a complicated way. Consider
first the model of non-interacting coherent electrons. A coherent
1D electron wave in a disordered 1D wire of infinite length is
exponentially localized by an arbitrary weak disorder
\cite{Mott-61,Borland-61} (it does not exhibit Anderson transition
\cite{Mott-90}). The resistance of a finite 1D wire should
therefore increase exponentially with the wire length. In fact,
the resistance wildly fluctuates from wire to wire in an ensemble
of macroscopically identical wires (because disorder in each wire
is microscopically different) and what increases exponentially is
the average resistance. \cite{Landauer-70,Anderson-80} A full
distribution of resistances in the ensemble of weakly disordered
1D wires is given by the Dorokhov-Mello-Pereyra-Kumar (DMPK)
equation -- an important result of the scaling theory of
localization. \cite{Beenakker-97}

How are these localization effects modified by many-body
interactions? As discussed below, this is an open question
especially for a single-channel 1D wire.

Authors of Ref.~\onlinecite{Matveev} considered tunneling of the
interacting 1D electron gas through a single impurity and examined
how the many-body interaction renormalizes single-particle
tunneling. They found at zero temperature, that the 1D electron at
the Fermi level is perfectly reflected by Friedel oscillations of
the Hartree-Fock many-electron potential around the impurity.
Thus, the impurity is impenetrable and the zero temperature
conductance is zero.

However, for such perfect reflection an infinitely long tail of
Friedel oscillations is needed. Therefore, a single impurity in
the wire of finite length still exhibits a nonzero (albeit
reduced) penetrability. \cite{Matveev} Many impurities distributed
at random along the finite-length wire can thus be expected to
cause a similar (exponential) localization as in the case of
non-interacting gas.

So far we have discussed coherent transport. Indeed, localization
and Friedel oscillations are due to the interference of incident
and reflected electron waves, for which electron coherence is
needed. However, the electron-electron (e-e) interaction in
general causes also the phase-breaking e-e collisions and thus
acts against interference. Recent theories of the e-e interaction
mediated phase breaking \cite{Altshuler-85,Zaikin-00} are
applicable to a quasi-1D (multi-channel) wire, not to the 1D
quantum wire with a single conducting channel. The same can be
said about experiments \cite{Mohanty-00,Gershenson-99} in which
the phase breaking is detected from magnetotransport in weakly
localized regime.

The e-e interaction mediated phase breaking in the 1D quantum wire
should in principle be tractable within the Landauer conductance
formulation (see the discussion in Sect.~IV). Various
Landauer-type formulations were used to analyze the effect of
point-like phase breakers placed at fixed positions in a random
chain of elastic scatterers (e.g. Refs.~\onlinecite{Buttiker-86}
and \onlinecite{Mello-00}), but the phase breaking by e-e
interaction was not considered.

Authors of Refs. \onlinecite{Dorokhov} and
\onlinecite{Shepelyansky-94} considered two interacting electrons
moving in disorder and found that the electron localization length
is enhanced in comparison with the single-electron situation.
Obviously, the phase breaking is not identified in this approach
as the interference (exponential localization) still exists. The
phase breaking which destroys exponential localization can perhaps
arise in the analysis involving many interacting 1D particles: The
many-particle interaction redistributes energy in a clean 1D
electron system, \cite{Sirenko-94,Mosko-94} so perhaps this
inelastic process persists also in a disordered 1D system.

Such many-particle analysis is a formidable problem, but a few
interesting questions can be addressed within a simple
phenomenological model developed in this work. In particular, as
already mentioned, coherent transport in a weakly disordered 1D
wire is described by the DMPK equation, \cite{Beenakker-97} which
determines the distribution of resistances in the ensemble of
macroscopically identical wires (eq.~\ref{e5}). We want to show
how the phase-breaking collisions, if any, modify the DMPK
distribution and the mean and typical resistance related to this
distribution.

We examine transport in a disordered 1D wire with a single
conducting channel, specifically in the ground subband of a GaAs
quantum wire. Our model is in essence a simple phenomenological
model of the e-e interaction mediated phase breaking, with the
electron coherence length treated as a parameter. In the model the
phase-breaking collisions effectively break the wire into
independent segments, where each segment is a series resistor with
coherent electron motion. We model the segmentation as a
stochastic process with a Poisson distribution of phase-breaking
scattering times and evaluate the resistance of each segment
microscopically from the Landauer formula.

The wire resistance as a function of the wire length $L$,
coherence length $L_{\phi}$, and localisation length $\xi$ is
traced from the coherent regime at $L\!<\!L_{\phi}$ up to the
incoherent one at $L\!\gg\!L_{\phi}$. We find the following
results.

In the coherent regime the resistance fluctuates from wire to wire
in accord with the DMPK resistance distribution, the typical
resistance increases as $\exp(L/\xi)$, and the mean resistance
increases as $\exp(2L/\xi)$. This is in agreement with the
universal scaling theory.
\cite{Anderson-80,Beenakker-97,Shapiro-87}

As $L$ exceeds $L_{\phi}$, decoherence suppresses the resistance
fluctuations and narrows the resistance distribution. As a result,
at $L\!\gg\!L_{\phi}$ the mean resistance increases as $\beta L-c$
and the typical resistance as $\beta L- c^\prime $, where $\beta$
is the wire resistivity, $c$ is a constant shift due to the
decoherence near the source electrode, and $c^\prime \gg c$ is the
shift related to the resistance self-averaging in a single wire.

In the GaAs quantum wire the backscattering by disorder can be
strong. \cite{Mosko-99} We therefore also examine the effect of
strong disorder on coherent transport. We find deviations from
universal scaling which differ from those reported for the
Anderson model. \cite{Slevin-90}

In Sect.~II we review the universal scaling theory and introduce
our microscopic model of coherent 1D transport. In Sect.~III the
effect of strong disorder on coherent transport is discussed and
deviations from universal scaling are reported. In Sect.~IV we
describe our decoherence model. Results of the decoherence model
are presented in Sect.~V. Discussion and conclusions are given in
Sect.~VI.

\section{Coherent 1D transport}

Electron transport in a disordered 1D wire with a single
conducting channel is described in a simple way, if one assumes
coherent transport of non-interacting quasi-particles. The wire
resistance $\rho$ (in units $h/2e^2$) reads
\cite{Landauer-70,Landauer-90,Datta-95}

\noindent
\begin{equation} \label{e1}
\rho=\frac{R(\varepsilon_F)}{T(\varepsilon_F)},
\end{equation}

\noindent
where $R$ and $T$ are the reflection and transmission
coefficients describing the electron tunneling through disorder at
Fermi energy. If disorder is specified by a random 1D potential
$V(x)$, then $R$ and $T$ can be obtained as

\noindent
\begin{equation} \label{e2}
R={|r_k|}^2, \qquad T={|t_k|}^2
\end{equation}

\noindent
by solving the tunneling problem

\noindent
\begin{equation} \label{e3}
\left[-\frac{\hbar^2}{2m}\
\frac{{\rm d}^2}{{\rm d}x^2}+V(x)\right] \Psi_k(x)={\cal
E}\Psi_k(x),
\end{equation}

\noindent
\begin{equation} \label{e4a}
\Psi_k\left(x\rightarrow0\right)=e^{ikx}+r_k e^{-ikx},
\end{equation}

\noindent
\begin{equation} \label{e4b}
\Psi_k\left(x\rightarrow L\right)=t_k e^{ikx},
\end{equation}

\noindent
where $\Psi_k(x)$ is the 1D electron wave function,
${\cal E}=\hbar^2k^2/2m$ is the electron energy, $m$ is the
effective mass, $L$ is the wire length (the source contact is
assumed at $x=0$, the drain contact at $x=L$), and $r_k$ and $t_k$
are the reflection and transmission amplitudes.

The wire resistance (\ref{e1}) however depends on microscopic
details of disorder and wildly fluctuates from wire to wire in an
ensemble of macroscopically identical wires. Therefore, instead of
the resistance $\rho$ of a single disordered wire the resistance
distribution $p(\rho)$ is meaningful. For weak disorder $p(\rho)$
is given by the DMPK equation. \cite{Beenakker-97} For a
single-channel wire the DMPK equation reads \cite{Melnikov}

\noindent
\begin{equation} \label{e5}
\xi \frac{\partial}{\partial L} \, p(\rho,L) =
\frac{\partial}{\partial\rho}
\left[
  \left(\rho^2+\rho\right) \frac{\partial}{\partial\rho} \, p(\rho,L)
\right] ,
\end{equation}

\noindent
where $\xi$ is the electron localisation length. Now we
review those properties of eq.~\ref{e5}, to which we refer later
on.

From~(\ref{e5}) one easy obtains the mean resistance

\noindent
\begin{equation} \label{e7}
\bar\rho(L) \equiv \int \limits _0 ^\infty d\rho\ \rho\
p(\rho,L)=\frac{1}{2} \left[\exp(2L/\xi)-1\right],
\end{equation}

\noindent
and also the mean square

\noindent
\begin{multline} \label{e8}
\bar{\rho^2} (L) \equiv \int \limits _0 ^\infty d\rho\
\rho^2 p(\rho,L)
\\
= \frac{1}{12} \left[ 2 \exp(6L/\xi) - 6 \exp(2L/\xi)+4 \right] .
\end{multline}

\noindent One sees that the dispersion
${(\bar{\rho^2}-{\bar\rho}^2)}^{1/2}/ \bar\rho\simeq\exp(L/\xi)$
for $L/\xi\gg1$, which means that $\bar\rho$ is not representative
of the ensemble. Anderson et al. \cite{Anderson-80} proposed to
average the variable $f=\ln(1+\rho)$. From eq.~(\ref{e5}) one
finds

\noindent
\begin{equation}
\bar{f} \equiv \int \limits _0 ^\infty d\rho \ln(1+\rho)
p(\rho,L)=\frac{L}{\xi}
\end{equation}

\noindent
and in a similar way (but in the limit $L/\xi\gg1$) also

\noindent
\begin{equation}
\Delta^2 \equiv \bar{f^{^2}}-{\bar{f}}^{^2}= \frac{2L}{\xi} .
\end{equation}

\noindent
Now the dispersion $\Delta/\bar{f}={(L/2\xi)}^{-1/2}$
decreases with $L/\xi$, so $\bar{f}$ is representative of the
ensemble. If one defines $p(\rho,L)={(1+\rho)}^{-1}{\cal
P}(\ln(1+\rho),L)$, eq.~(\ref{e5}) readily gives

\noindent
\begin{equation} \label{e11}
{\cal P}(\ln\rho,L) = \frac{1}{\sqrt{2\pi\Delta^2}}
\exp\left[-\frac{{(\ln\rho-\bar{f})}^2}{2\Delta^2}\right]
\end{equation}

\noindent for $\rho\gg1$ (the case whenever $L/\xi\gg1$).
Expression (\ref{e11}) is a Gauss distribution centered at
$\bar{f}=L/\xi$ with a spread $\Delta^2=2\bar{f}=2L/\xi$. Thus,
$p(\rho,L)$ is a log-normal distribution with a Gaussian-shaped
bulk parametrized by $\bar{f}=L/\xi$. The moments of the
distribution $p(\rho,L)$ [eqs.~\ref{e7},~\ref{e8}] are dominated
by the $1/\rho$ tail. One can introduce \cite{Anderson-80} the
typical resistance $\rho_t$ by definition $\ln(1+\rho_t)=\bar{f}$.
For coherent transport

\noindent
\begin{equation} \label{e12}
\rho_t(L)=\exp(L/\xi)-1 ,
\end{equation}

\noindent
since $\bar{f}=L/\xi$. Obviously, the distribution
(\ref{e11}) is peaked at $\rho=\rho_t$, i.e., $\rho_t$ is
insensitive to the tail of $p(\rho,L)$.

\begin{figure}[t]
\noindent
\centerline{\includegraphics[clip,width=0.85\columnwidth]
{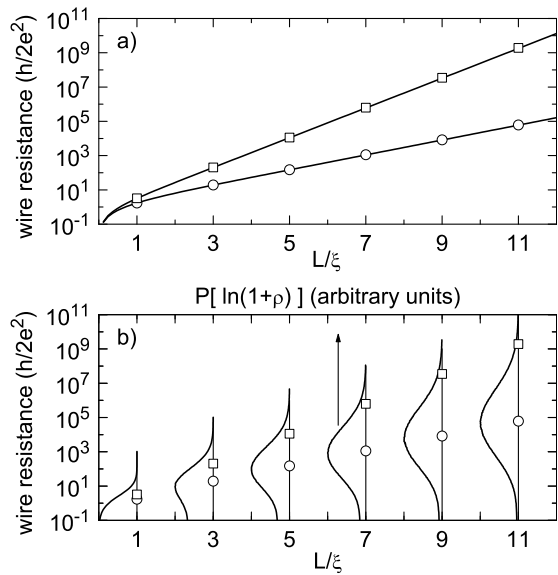}} \caption{ (a) Mean resistance (squares) and typical
resistance (circles) versus the scaling parameter $L/\xi$ for the
1D wire with weak disorder as described in the text. Squares and
circles are the microscopic model data, lines connecting these
data are graphic representation of the formulae (\ref{e7}) and
(\ref{e12}), respectively. (b) Distribution ${\mathcal
P}(\ln(1+\rho))$ versus $\rho$: microscopic results for the same
$L/\xi$ as the squares and circles. } \label{Fig1}
\end{figure}

Equations (\ref{e5})--(\ref{e12}) are macroscopic, their
fundamental feature is the universal scaling with a single
macroscopic parameter $L/\xi$. Now we describe our microscopic
model.

We consider disorder $V(x)=\sum _{i=1}^N \gamma\delta(x-x_i)$,
where $\gamma\delta(x-x_i)$ is the $\delta$-shaped impurity
potential of strength $\gamma$, $x_i$ is the $i$-th impurity
position (selected at random along the wire), and $N$ is the
number of impurities. We select $N$ at random from the Poisson
distribution

\noindent
\begin{equation} \label{e13}
{\mathcal G}(N)={(N_IL)}^Ne^{-N_IL}/N!
\end{equation}

\noindent
where $N_I$ is the linear impurity density. The
reflection coefficient of a single $\delta$-barrier is $R_I=
\Omega^2/(k_F^2+\Omega^2)$, where $\Omega=m\gamma/\hbar^2$. We fix
$k_F=7.9\times10^7$ m$^{-1}$ and $m=0.067m_0$, and we parameterize
the $\delta$-barrier by $R_I$.

We ignore the fluctuations of $R_I$, because they make the
presentation less transparent but all major results remain the
same (for weak disorder universal scaling holds independently on
the choice of disorder) or similar (for strong disorder).

The scheme of the microscopic modeling is simple. We select
disorder, solve eq.~(\ref{e3}) by the transfer matrix method,
\cite{transfermatrix} and obtain from eq.~(\ref{e1}) the
resistance of a single wire. We repeat this process for the
ensemble of wires and obtain the resistance distribution and mean
values.

Figure~\ref{Fig1} shows results of the microscopic model for $R_I=
0.01$ and $N_I=10^6$ m$^{-1}$. Since $R_I\ll1$ and
$N_I^{-1}\gg2\pi/k_F$, we can speak about the weak low-density
disorder. In such case eqs. (\ref{e5}-\ref{e12}) hold for
\cite{Anderson-80,Beenakker-97}

\noindent
\begin{equation} \label{e14}
\xi=l={(N_IR_I)}^{-1} ,
\end{equation}

\noindent
where $l$ is the classical elastic mean free path
($100$~$\mu$m in our case). If we set this value into
eqs.~(\ref{e7}) and~(\ref{e12}), they indeed reproduce the
microscopic data in Fig.~\ref{Fig1}a. We also see in
Fig.~\ref{Fig1}b, that with increasing $L/\xi$ the distribution
${\cal P}(\ln(1+\rho))$ tends to center around the typical
resistance, while the mean resistance is essentially out of the
distribution. This ${\cal P}(\ln(1+\rho))$ dependence can be
reproduced by the formula~(\ref{e11}) with $\xi=100$~$\mu$m.

In summary, microscopic model and the scaling theory give the same
results, as one expects for weak disorder. However, in the 1D GaAs
wire the backscattering from a single impurity can be quite strong
\cite{Mosko-99} ($R_I\approx0.1-0.9$). In the next section we show
that this can cause peculiar deviations from universal scaling.

\section{Deviations from universal scaling}

We still consider the chain of $N$ randomly-positioned identical
$\delta$-barriers at density as low as $N_I^{-1}\gg2\pi/k_F$. For
$N_I^{-1}\gg2\pi/k_F$ useful exact expressions hold. The mean
resistance reads

\noindent
\begin{equation} \label{e15}
\bar{\rho} (N) = \frac{1}{2} \left[
{\left(\frac{1+R_I}{1-R_I}\right)}^N - 1 \right],
\end{equation}

\noindent
as shown for the first time by Landauer.
\cite{Landauer-70} We rederive this expression in a more simple
way in the Appendix. In the Appendix we also derive the mean
square

\noindent
\begin{equation} \label{e16}
\bar{\rho^2}(N) = \frac{1}{12} \left[
2 {\left( 1+\frac{6R_I}{{(1-R_I)}^2} \right)}^N \right.
- 6 \left. {\left(\frac{1+R_I}{1-R_I}\right)}^N+4 \right]
\end{equation}

\noindent
and the typical resistance

\noindent
\begin{equation} \label{e17}
\rho_t(N)={\left(\frac{1}{1-R_I}\right)}^N-1.
\end{equation}

\noindent
Equations~(\ref{e15}--\ref{e17}) are exact for any
$R_I$, if $N_I^{-1}\gg2\pi/k_F$. Averaging them over the
distribution~(\ref{e13}) we include fluctuations of $N$. We obtain
equations

\noindent
\begin{equation} \label{e18}
\bar{\rho}(L) =
\frac{1}{2}\left[\exp\left(2N_I\frac{R_I}{1-R_I}L\right)-1 \right]
,
\end{equation}

\noindent
\begin{multline} \label{e19}
\bar{\rho^2}(L) =
\frac{1}{12}\left[2\exp\left(6N_I\frac{R_I}{{(1-R_I)}^2}L
\right) \right.
\\
- \left. 6 \exp \left( 2N_I\frac{R_I}{1-R_I} L \right)+4 \right] ,
\end{multline}

\noindent
\begin{equation} \label{e20}
\rho_t(L) = \exp \left[ N_I \ln \left( \frac{1}{1-R_I} \right) L
\right]-1 ,
\end{equation}

\noindent
which provide us with the exact dependence on $L$.

\begin{figure}[t]
\noindent
\centerline{\includegraphics[clip,width=0.85\columnwidth]
{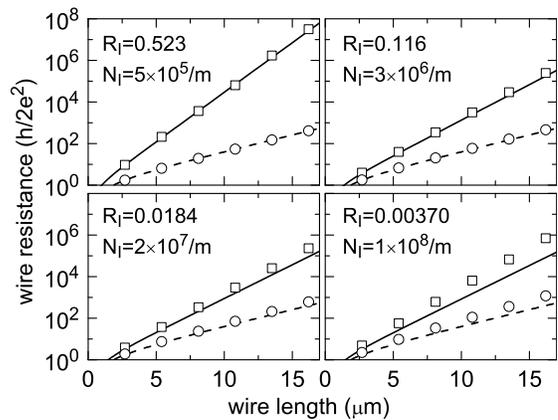}} \caption{ Mean resistance (squares, full lines) and
typical resistance (circles, dashed lines) versus the wire length
$L$. Squares and circles are the microscopic model results, full
lines and dashed lines are graphic representation of the formulae
(\ref{e18}) and (\ref{e20}), respectively. Parameters $N_I$ and
$R_I$ are varied in such way that the localization
length~(\ref{e21}) is the same ($\xi=2.7$~$\mu$m) for each figure.
One can see that the accuracy of eqs.~(\ref{e18}) and (\ref{e20})
deteriorates with increasing $N_I$. } \label{Fig7}
\end{figure}

Let us compare eqs.~(\ref{e18}), (\ref{e19}), and (\ref{e20}) with
the scaling theory equations. Comparing eq.~(\ref{e20}) with
eq.~(\ref{e12}) we get the localisation length

\noindent
\begin{equation} \label{e21}
\xi={\left[N_I\ln\left(\frac{1}{1-R_I}\right)\right]}^{-1}.
\end{equation}

\noindent
Comparing eq.~(\ref{e18}) with eq.~(\ref{e7}) we obtain
the length

\noindent
\begin{equation} \label{e22}
\xi_1={\left(N_I\frac{R_I}{1-R_I}\right)}^{-1},
\end{equation}

\noindent
where index $1$ is added to distinguish from $\xi$.
Finally, comparing eq.~(\ref{e19}) with eq.~(\ref{e8}) we again
find the length~(\ref{e22}) and additionally the length

\noindent
\begin{equation} \label{e23}
\xi_2={\left[N_I\frac{R_I}{{(1-R_I)}^2}\right]}^{-1}.
\end{equation}

A comment on the accuracy of the above equations. If
$N_I^{-1}\gg2\pi/k_F$, our microscopic model gives results, which
agree with eqs.~(\ref{e18})--(\ref{e20}) for any value of $R_I$.
Otherwise, a simple phase averaging used in the Appendix
[eqs.~(\ref{Apprho}),~(\ref{AppLandTyp1}),~(\ref{AppAllQuant})] is
no longer accurate and the accuracy of
eqs.~(\ref{e18})--(\ref{e20}) deteriorates as we demonstrate in
Fig.~\ref{Fig7}. In such case eqs.~(\ref{e21})--(\ref{e23}) are
not reliable as well and we have to obtain $\xi$, $\xi_1$, and
$\xi_2$ directly from the microscopic model. (This means that we
fit the microscopic model results by the formulae~(\ref{e12}),
(\ref{e7}), and (\ref{e8}) with a properly adjusted $\xi$,
$\xi_1$, and $\xi_2$, respectively. To obtain the localization
length $\xi$, we can also use the Lyapunov exponent analysis,
\cite{Markos} but we get the same result.) In this section we
restrict us to the limit $N_I^{-1}\gg2\pi/k_F$, because
eqs.~(\ref{e18}), (\ref{e19}), and (\ref{e20}) are useful and our
conclusions would remain similar also beyond the limit
$N_I^{-1}\gg2\pi/k_F$.

Equations~(\ref{e18}), (\ref{e19}), and (\ref{e20}) scale with
three parameters $L/\xi$, $L/\xi_1$, and $L/\xi_2$ and coincide
with eqs.~(\ref{e7}), (\ref{e8}), and (\ref{e12}) only in the
limit $R_I \ll 1$, when $\xi=\xi_1=\xi_2={(N_IR_I)}^{-1}$. Since
the moments (\ref{e18}) and (\ref{e19}) do not scale with the same
parameter as the typical resistance (\ref{e20}), this invokes that
also the bulk of the distribution, i.e., ${\cal P}(\ln(1+\rho))$,
does not scale with a single parameter.

\begin{figure}[t]
\noindent
\centerline{\includegraphics[clip,width=0.85\columnwidth]
{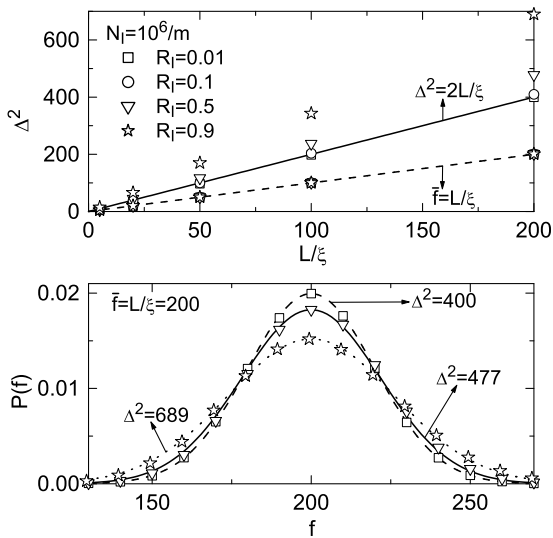}} \caption{ The top figure: Mean deviation $\Delta^2
\equiv \bar{f^{^2}}-{\bar{f}}^{^2}$ versus $L/\xi$ ($\xi$ is given
by eq.~(\ref{e21})) for various $R_I$. The microscopic model
(symbols) is compared with the scaling theory result
$\Delta^2=2L/\xi$ (full line). Also is compared $\bar{f}=L/\xi$
(dashed line) with $\bar{f}$ obtained from the microscopic model
(symbols coinciding with the dashed line). The bottom figure:
Distribution ${\mathcal P}(f)$ for $L/\xi=200$. Symbols show the
microscopic results for various $R_I$. Lines connecting points are
the Gauss distributions ~(\ref{e11}) with $\Delta^2$ as indicated,
the indicated data are microscopic data from the top figure.}
\label{Fig2}
\end{figure}

Our microscopic model indeed shows that the distribution ${\cal
P}(\ln(1+\rho))$ scales with two parameters. This is demonstrated
in Fig.~\ref{Fig2} for various $R_I$ and for $N_I=10^6$ m$^{-1}$.
The top figure shows that for large $R_I$ the mean deviation
$\Delta^2$ does not scale as $2L/\xi$. It can be seen that in
general $\Delta^2>2L/\xi$. Further, as shown in the bottom figure,
the microscopically calculated ${\cal P}(\ln(1+\rho))$ coincides
with the Gauss distribution~(\ref{e11}) centered at
$\bar{f}=L/\xi$, but its spread $\Delta^2$ is no longer $2L/\xi$.

\begin{figure}[t]
\noindent
\centerline{\includegraphics[clip,width=0.85\columnwidth]
{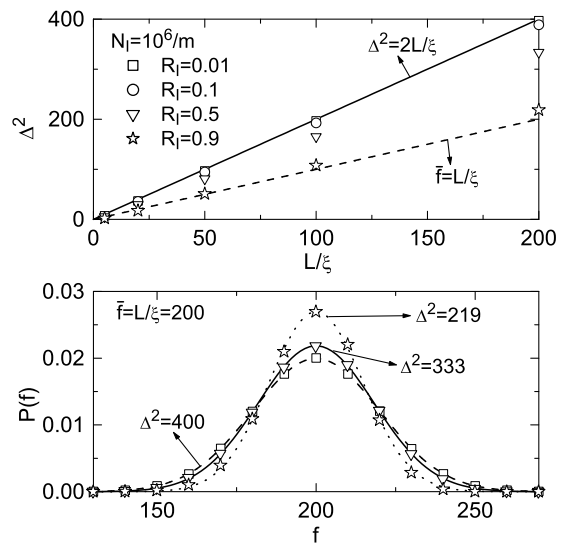}} \caption{ The same calculation as in Fig.~\ref{Fig2},
but with $N$ fixed to its mean value $N_IL$. Also symbols have the
same meaning as in Fig.~\ref{Fig2}. } \label{Fig3}
\end{figure}

Two-parametric scaling has already been reported for the 1D
Anderson model, \cite{Slevin-90} in which disorder is due to the
equidistant barriers with a barrier strength fluctuating at
random. It has been found \cite{Slevin-90} that $\Delta^2<2L/\xi$
rather than $\Delta^2>2L/\xi$. This difference is due to the fact
that in our model $N$ fluctuates from wire to wire, as is the case
for impurity disorder in real samples. If we fix $N$ in each wire
to its mean value $N_IL$, we also obtain $\Delta^2<2L/\xi$ (see
Fig.~\ref{Fig3}). The two-parameter scaling (Fig.~\ref{Fig3}) is
then similar to that one in the Anderson model. \cite{Slevin-90}

Why the fluctuating $N$ enhances $\Delta^2$ in comparison with the
case $N=N_IL$? In the ensemble with fluctuating $N$ the wires with
$N>N_IL$ ($N<N_IL$) typically contribute into the resistance
distribution by larger (smaller) resistance values than the wires
with $N=N_IL$. This broadens the distribution.

Why the fluctuating $N$ does not affect $\Delta^2$ in the limit of
small $R_I$? The dispersion of $N$ is $1/\sqrt{N_IL}$. As $R_I$
approaches zero, $L$ approaches infinity in order to keep a
constant $L/\xi$. The dispersion of $N$ thus becomes negligible,
so one obtains $\Delta^2=2L/\xi$ for the fluctuating as well as
fixed $N$ (c.f. Figs.~\ref{Fig2} and~\ref{Fig3}).

\section{Decoherence model}

Now we assume that the electron motion along the wire is perturbed
by inelastic collisions. An electron will move coherently with a
certain energy for a while, suffer an inelastic collision which
will transfer it to another energy, move coherently with the new
energy, suffer another inelastic collision and so on. Thus,
electron motion across disorder is no longer elastic and the final
states after inelastic collisions are in general affected by the
blocking effect of Pauli principle. In this situation the
applicability of the Landauer expression~(\ref{e1}) is
questionable.

However, as pointed out in Ref.~\onlinecite{Datta-95},
expression~(\ref{e1}) is still applicable if one assumes that
there is no net ``vertical'' flow of electrons: every electron
that scatters out of its energy ${\cal E}_1$ into another energy
${\cal E}_2$ is balanced by another electron that scatters out of
${\cal E}_2$ into ${\cal E}_1$. This effectively means that an
electron with a given energy moves coherently for a while, suffers
a phase-breaking collision, continues to move coherently with the
same energy, suffers another phase-breaking collision and so on.
This ``effective'' electron feels the phase-breaking collisions
like elastic events and the effect of Pauli blocking disappears.
Transport thus still proceeds via independent energy channels and
can be described in terms of the transmission and reflection as
does equation~(\ref{e1}). However, the transmission of electrons
from one contact to the other is now characterised by repeated
phase breaking rather than by a single coherent process.

To account for the repeated phase breaking in a simple way we
further assume that each phase-breaking collision randomizes the
electron phase completely. \cite{complete dephasing} In such case
it is expected \cite{Kramer-89} and can be shown rigorously,
\cite{Buttiker-86} that the phase-breaking events break the wire
into segments where each segment is an independent series resistor
with coherent electronic resistance. The resistance of each
segment can thus be evaluated from the Landauer resistance formula
and the resistance of the wire is simply a sum of the resistances
of all segments.

We recall that our model relies on the assumption that there is no
net ``vertical'' flow. This assumption is exact for the e-e
interaction. Indeed, if only the e-e interaction is operative,
then there is no energy exchange between the electrons and crystal
lattice, so any non-zero ``vertical'' flow would distort the
electron energy distribution. In fact the e-e interaction
maintains a thermalised distribution, \cite{Mosko-94} i.e., there
is no such distortion. Moreover, the e-e interaction conserves the
total momentum of the 1D gas and does not cause any
momentum-relaxation related resistance. All these features are
inherent to our model, so we believe that we have a reasonable
phenomenological model of the e-e interaction mediated phase
breaking.

Now we give details of our model. Consider a single wire with
specified disorder. Assume that an electron at the Fermi level
undergoes $N$ inelastic collisions when traversing the wire length
$L$. Denote the time elapsed between the $n-1$ and $n$ collisions
as $\tau_{n-1, n}$ and the positions of these collisions along the
wire as $x_{n-1}$ and $x_n$. Due to disorder the electron motion
is diffusive, therefore

\noindent
\begin{equation} \label{diffus}
{(x_n-x_{n-1})}^2=D\tau_{n-1,n} ,
\end{equation}

\noindent
where $D$ is the elastic diffusion coefficient. The wire
is divided into segments $x_n-x_{n-1}$, where each segment is a
series resistor with coherent resistance $\rho(x_n-x_{n-1})$. We
evaluate $\rho(x_n-x_{n-1})$ from eq.~(\ref{e1}) by solving
eq.~(\ref{e3}) for the boundary conditions (\ref{e4a}) and
(\ref{e4b}) applied at the boundaries $x_{n-1}$ and $x_n$. The
resistance of the segmented wire reads $\rho(\tau_{0,1}) +
\rho(\tau_{1,2}) + \dots + \rho(\tau_{N-1,N}) + \rho(\tau_{N,L})$,
where the variable $\tau_{n-1,n}$ is used instead of $x_n-x_{n-1}$
and the positions at the beginning and end of the wire are denoted
as $0$ and $L$.

Of course, when transmitted from the source to the drain, each
Fermi electron experiences a different random configuration of
segments. Therefore, the resistance of the segmented wire has to
be averaged over all possible configurations of segments, taken
with a proper weight.

For this purpose, denote the inelastic scattering rate as
$1/\tau_{in}$. Let $P(\tau)$ be the probability that the electron
moves without inelastic collision for time $\tau$. The probability
that one inelastic collision occurs in time $(\tau, \tau+d\tau)$
is given by $P(\tau) d\tau / \tau_{\text{in}}
=P(\tau)-P(\tau+d\tau)$, which gives

\noindent
\begin{equation} \label{Probab}
P\left( \tau \right) \frac{d\tau} {\tau_{\text{in}}} = \exp \left(
-\frac{\tau} {\tau_{\text{in}}} \right)
\frac{d\tau}{\tau_{\text{in}}} .
\end{equation}

\noindent
For $N=0$ the weighted resistance is simply

\noindent
\begin{equation} \label{R0}
\rho_{N=0} \left(L \right) = \rho \left( \tau_{0,L} \right) \exp
\left( - \frac{\tau_{0,L}} {\tau_{\text{in}}} \right) ,
\end{equation}

\noindent
where $\tau_{0,L}=L^2/D$ is the elastic diffusion time
from the source up to the drain. Similarly, for $N=1$ one finds

\noindent
\begin{multline} \label{R1}
\rho_{1} \left(L \right) = \int \limits _{0} ^{\tau_{0,L}} \frac
{d\tau_{0,1}} {\tau_{\text{in}}} \exp \left( - \frac{\tau_{0,1}}
{\tau_{\text{in}}} \right) \exp \left( - \frac{\tau_{1,L}}
{\tau_{\text{in}}} \right)
\\ \times
\left[ \rho \left( \tau_{0,1} \right) + \rho \left( \tau_{1,L}
\right) \right] ,
\end{multline}

\noindent where $L = \sqrt{D} \, \left( \sqrt{\tau_{0,1}} +
\sqrt{\tau_{1,L}} \right)$, because the segments must add to give
the total wire length $L$. Analogously,

\noindent
\begin{multline} \label{R2}
\rho_{2} \left(L \right) = \int \limits _{0} ^{\tau_{0,L}} \frac
{d\tau_{0,1}} {\tau_{\text{in}}} \int \limits _{0}
^{\tau_{1,2}^{\text{max}}}
\frac {d\tau_{1,2}} {\tau_{\text{in}}}
\\ \times
\exp \left( - \frac{\tau_{0,1}} {\tau_{\text{in}}} \right) \exp
\left( - \frac{\tau_{1,2}} {\tau_{\text{in}}} \right) \exp \left(
- \frac{\tau_{2,L}} {\tau_{\text{in}}} \right)
\\ \times
\left[ \rho \left( \tau_{0,1} \right) + \rho \left( \tau_{1,2}
\right) + \rho \left( \tau_{2,L} \right) \right] ,
\end{multline}

\noindent
where $L = \sqrt{D} \, ( \sqrt{\tau_{0,1}} +
\sqrt{\tau_{1,2}} + \sqrt{\tau_{2,L}} )$ and
$\tau_{1,2}^{\text{max}} = (\sqrt{\tau_{0,L}} - \sqrt{\tau_{0,1}}
)^2$. It is easy to generalize $\rho_2 (L)$ to $\rho_N (L)$ and to
express the averaged resistance as

\noindent
\begin{multline} \label{RInfty}
\rho \left(L \right) = \sum \limits _{N=0} ^{\infty} \rho_N \left(
L \right) =\sum \limits _{N=0} ^{\infty} \int \limits _{0}
^{\tau_{0,L}} \frac {d\tau_{0,1}} {\tau_{\text{in}}} \int \limits
_{0} ^{\tau_{1,2}^{\text{max}}} \frac {d\tau_{1,2}}
{\tau_{\text{in}}} \, \dots
\\ \times
\int \limits _{0} ^{\tau_{N-1,N}^{\text{max}}} \frac
{d\tau_{N-1,N}} {\tau_{\text{in}}} \exp \left( - \frac{\tau_{0,1}}
{\tau_{\text{in}}} \right) \exp \left( - \frac{\tau_{1,2}}
{\tau_{\text{in}}} \right) \, \dots
\\ \times
\exp \left( -
\frac{\tau_{N-1,N}} {\tau_{\text{in}}} \right) \exp \left( -
\frac{\tau_{N,L}} {\tau_{\text{in}}} \right)
\\ \times
\left[ \rho \left( \tau_{0,1} \right) + \rho \left( \tau_{1,2}
\right) + \dots + \rho \left( \tau_{N-1,N} \right) + \rho \left(
\tau_{N,L} \right) \right] ,
\end{multline}

\noindent
where $L/\sqrt{D} = \sqrt{\tau_{0,1}} +
\sqrt{\tau_{1,2}} + \dots + \sqrt{\tau_{N-1,N}} +
\sqrt{\tau_{N,L}}$ and $\tau_{n-1,n}^{\text{max}} = \left(
\sqrt{\tau_{0,L}} - \sqrt{\tau_{0,1}} - \sqrt{\tau_{1,2}} - \dots
- \sqrt{\tau_{n-2,n-1}} \right)^2$.

We recall that eq.~(\ref{RInfty}) gives the resistance of a wire
with specific configuration of disorder. Evaluating
eq.~(\ref{RInfty}) for the ensemble of wires with different
configurations of disorder we can obtain numerically the
resistance distribution and the mean and typical resistance.

One has to be careful when evaluating the typical resistance. Our
derivation of eq.~(\ref{RInfty}) is motivated by
Ref.~\onlinecite{Farrell-90}, which uses similar considerations to
derive the typical resistance. Equation (3.7) in
Ref.~\onlinecite{Farrell-90} should coincide with our
result~(\ref{RInfty}), but it does not: the limits
$\tau_{n-1,n}^{\text{max}}$ are different. Another problem,
physical rather than technical, is the following. It is tempting
to follow Ref.~\onlinecite{Farrell-90} and to set for each segment
in eq.~(\ref{RInfty}) the typical coherent resistance

\noindent
\begin{equation} \label{resist2}
\rho(\tau_{n-1,n}) = \exp \left( \frac{\sqrt{D\, \tau_{n-1,n}}}
{\xi} \right)-1 .
\end{equation}

\noindent
However, the typical resistance of the segmented wire is
not a sum of typical resistances of the individual segments. A
correct approach is to obtain the ensemble average $\bar{f}$ of
the quantity $f=\ln(1+\rho(L))$, where $\rho(L)$ is given by
eq.~(\ref{RInfty}), and to extract the typical resistance from
definition $\ln(1+\rho_t)=\bar{f}$.

On the other hand, the mean resistance of the segmented wire as a
direct ensemble average of eq.~(\ref{RInfty}) is a sum of mean
resistances of individual segments. Thus, if we set for each
segment the mean coherent resistance

\noindent
\begin{equation} \label{meanresist2}
\rho(\tau_{n-1,n}) = \frac{1}{2} \left[ \exp \left(
\frac{2\sqrt{D\, \tau_{n-1,n}}} {\xi_1} \right)-1\right] ,
\end{equation}

\noindent
then eq.~(\ref{RInfty}) expresses the mean resistance of
the wire.

We have derived eq.~(\ref{RInfty}) to provide insight and to give
a formal expression for the wire resistance. Now we describe
another averaging procedure, a Monte Carlo approach, which does
not provide any formulae, but is much easier to implement and
gives the same numerical results.

We start by specifying disorder in a given wire. The segmentation
of the wire is then simulated as a Monte Carlo process, in which
the time $\tau$ between two inelastic collisions is selected at
random from equation

\noindent
\begin{equation} \label{Rand}
r= \int \limits _{0} ^{\tau} \frac {d\tau} {\tau_{\text{in}}} \exp
\left(- \frac {\tau} {\tau_{\text{in}}} \right) =1-\exp \left(-
\frac {\tau} {\tau_{\text{in}}} \right) ,
\end{equation}

\noindent
where $r$ is a random number between $0$ and $1$. We
generate the times $\tau_{0,1}$, $\tau_{1,2}$, $\tau_{2,3},\dots$.
After each time generation we determine the total segmented
length, say $\sqrt{D\tau_{0,1}}+\sqrt{D\tau_{1,2}}+
\sqrt{D\tau_{2,3}}$, and we check whether it does not exceed $L$.
If it does not, we generate $\tau_{3,4}$. If it does, we obtain
the length of the rest, $\sqrt{D\tau_{2,L}} = L - \left(
\sqrt{D\tau_{0,1}} + \sqrt{D\tau_{1,2}} \right)$. Thus, the
position and length of each segment between the source and drain
are known. We evaluate the coherent resistance of each segment and
sum the resistances of all segments to obtain the total resistance
of the segmented wire. We repeat this segmentation process for the
same wire many times and average the wire resistance over all
configurations of segments (numerical results are the same as
gives eq.~(\ref{RInfty})). We apply the described procedure to the
ensemble of wires with various configurations of disorder and
obtain the resistance distribution and ensemble-averaged
quantities.

\begin{figure}[t]
\noindent
\centerline{\includegraphics[clip,width=0.85\columnwidth]
{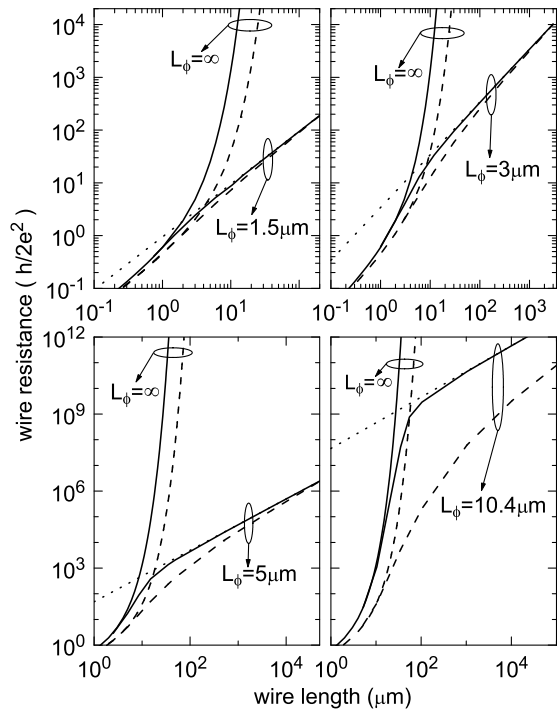}} \caption{ The mean resistance (solid lines) and
typical resistance (dashed lines) of the 1D wire as functions of
the wire length $L$ for various coherence lengths $L_{\phi}$. The
coherent resistance ($L_{\phi} = \infty$) is shown for comparison.
The dotted line shows the linear $\beta L$ slope (see the text) to
which the mean and typical resistances converge at large $L$. In
these calculations disorder with parameters $N_I=3\times 10^6$
m$^{-1}$ and $R_I=0.116$ was used, for these parameters the
microscopic approach of Sect.~III gives $\xi=2.65$~$\mu$m,
$\xi_1=2.47$~$\mu$m, and $\xi_2=2.15$~$\mu$m. } \label{Fig4}
\end{figure}

A comment about the parameters of our decoherence model.
Coefficient $D$ describes coherent diffusion across disorder.
Since the coherent resistance fluctuates from segment to segment,
also $D$ should be evaluated for each segment separately. For
simplicity, we assume in each segment the same $D$. This allows us
to substitute the variable $\tau$ by the variable $s
=\sqrt{D\tau}$ in all formulae of this section and we get rid of
$D$. The only parameter is thus $\tau_{in}$. Since the electron
coherence length is $L_{\phi}=\sqrt{D\tau_{in}}$, in what follows
we refer to $L_{\phi}$ rather then to $\tau_{in}$.

At low enough temperatures the electron-phonon interaction is
suppressed, which we assume. In such case $L_{\phi}$ is governed
by e-e interactions and the calculation of $L_{\phi}$ for a
disordered single-moded 1D wire is a formidable task. We therefore
consider $L_{\phi}$ as a parameter.

\section{Results of decoherence model}

In Fig.~\ref{Fig4} we show the mean and typical resistances in
presence of decoherence for various values of $L_{\phi}$ and for
disorder as specified in the figure caption. Disorder gives the
localization length $\xi=2.65$~$\mu$m, roughly this value is
typical say for a V-groove GaAs quantum wire. \cite{Kaufman-99} We
vary $L_{\phi}$ from smaller to larger values than $\xi$, since at
low temperatures crossover of this kind is expected.

Figure~\ref{Fig4} also shows the mean and typical resistance in
the coherent regime ($L_{\phi}=\infty$). For $L<L_{\phi}$ the
resistance with decoherence and the coherent resistance coincide
as one expects. For $L>L_{\phi}$ both the mean and typical
resistance are driven by decoherence towards a linear dependence
on $L$. The slope of this linear dependence is shown in a dotted
line, note that the mean resistance reaches the linear slope at
much smaller $L$ than does the typical resistance. Now we discuss
these results in detail.

\subsection{Resistivity and mean resistance}

The slope of the linear dependence in Fig.~\ref{Fig4} gives the
wire resistivity, $\beta$. For an infinite wire we expect

\noindent
\begin{equation} \label{beta}
\beta = \frac{
  \int \limits _0 ^\infty ds \, g \left(s\right) \rho\left(s\right)
} {
  \int \limits _0 ^\infty ds \, g \left(s\right) s
} ,
\end{equation}

\noindent
where $\rho(s)=0.5 \left[ \exp(2s/\xi_1)-1 \right]$, $s$
is the length of the segment, and the $s$--distribution for the
(infinite) wire is

\noindent
\begin{equation} \label{PDelta}
g \left(s \right) = \, \frac{2s}{L_{\phi}^2} \,
e^{-\frac{s^2}{L_{\phi}^2}} ,
\end{equation}

\noindent as follows from eq.~(\ref{Probab}) for $s
=\sqrt{D\tau}$. Equation (\ref{beta}) gives \cite{Blencowe}

\noindent
\begin{equation} \label{Blenc}
\beta = \frac{1}{\xi_1}\, \exp{\left( \frac{L_{\phi}^2}{\xi_1^2}
\right)} \left[ 1 + \text{erf} \left( \frac{L_{\phi}}{\xi_1}
\right) \right]
\end{equation}

\noindent and the dotted line in Fig.~\ref{Fig4} shows the
dependence $\beta L$. Clearly, the slope of the dotted line
coincides with the linear slope to which the decoherence drives
the mean and typical resistance. Thus, eq.~(\ref{Blenc}) correctly
expresses the wire resistivity. We have not attempted to do so,
but eq.~(\ref{Blenc}) should be derivable directly from
eq.~(\ref{RInfty}).

\begin{figure}[t]
\noindent
\centerline{\includegraphics[clip,width=0.85\columnwidth]
{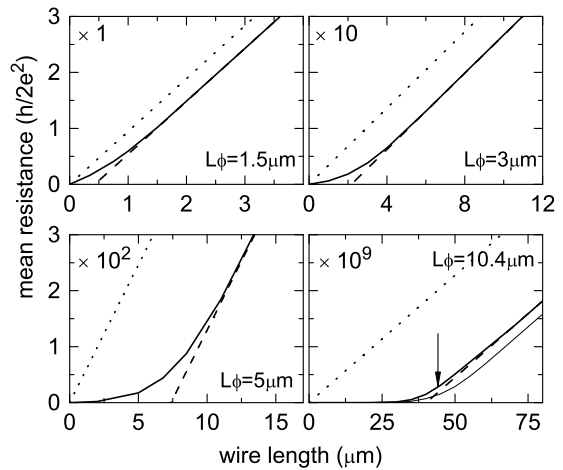}}
\caption{The mean resistance from Fig.~\ref{Fig4}
shown in a linear scale: the dotted line shows the dependence
$\beta L$, the full line is the mean resistance resulting from the
decoherence model, the dashed line is the shifted $\beta L$
dependence to which the mean resistance actually converges. The
resistance data should be multiplied by a factor $1$, $10$,
$10^2$, or $10^9$, as indicated. The arrow and thin full line are
discussed in the text. } \label{Fig5}
\end{figure}

The log-scale in Fig.~\ref{Fig4} obscures an interesting effect.
In Fig.~\ref{Fig5} we present the mean resistance from
Fig.~\ref{Fig4} in a linear scale. Clearly, the ohmic dependence
to which the decoherence drives the mean resistance is not $\beta
L$, but $\beta L-c$, where $c>0$ is a constant shift. The shift
increases with $L_{\phi}$ and diminishes for $L_{\phi}=0$, it
should be observed in the 1D wires long enough to exhibit the
ohmic resistance. As discussed below, this shift is a ``memory
effect'' reflecting the decoherence near the source contact.

To give insight we approximate the exact mean resistance (the
ensemble-averaged equation~(\ref{RInfty})) as

\noindent
\begin{equation} \label{rhoL}
\bar \rho (L) \approx \frac{1}{2}\, ( e^{ \frac{2L} {\xi_1}}-1 ) \
e^{- \frac{L^2}{L_{\phi}^2}} + (1-e^{- \frac{L^2}{L_{\phi}^2}})
 \int \limits _0 ^L dx \beta(x) ,
\end{equation}

\noindent
where $x$ is the distance from the source electrode and

\noindent
\begin{equation} \label{betax}
\beta (x) = \frac{ \int \limits _0 ^x ds \, g \left(s \right)
\frac{1}{2}\, ( e^{ \frac{2s} {\xi_1}}-1 ) } {
  \int \limits _0 ^x ds \, s \, g \left(s \right)
}
\end{equation}

\noindent
is the resistivity~(\ref{beta}) approximated for a
finite wire.  The first term on the right hand side of
eq.~(\ref{rhoL}) coincides with the first term of the
ensemble-averaged sum~(\ref{RInfty}) and the term $\int_{0}^L dx
\beta(x)$ approximates the rest of the sum (see below). The first
term is the contribution from coherent transmission while the term
$\int_{0}^L dx \beta(x)$ describes the contribution from
incoherent transmission. The weight $1-\exp{(-L^2/L_{\phi}^2)}$ is
the probability that the transmission is not coherent, it tailors
both terms at $L \approx L_{\phi}$.

Equation~(\ref{betax}) accounts for the fact that decoherence in
the wire increment $(0,x)$ does not involve segments larger than
$x$, but it still relies on eq.~(\ref{PDelta}), which strictly holds
only for the infinite wire. \cite{commentPdelta} The question is
now how good is
eq.~(\ref{rhoL}). In Fig.~\ref{Fig5} the dependence~(\ref{rhoL})
is shown in the thin full line for $L_{\phi} = 10.4$~$\mu$m. It
reasonably fits the thick full line, with increasing
$L_{\phi}/\xi_1$ (not shown) the fit is even better.

Equation~(\ref{rhoL}) thus explains the shift $c$ in the limit
$L_{\phi} \gg \xi_1$. It that limit $\beta (x) \ll \beta$ for $x$
smaller than a certain distance $L_0$, while for $x>L_0$ $\beta
(x)\simeq \beta$. Indeed, the function $g \left(s\right)
\rho\left(s\right)$ is peaked at

\noindent
\begin{equation} \label{L0}
L_0 \equiv \frac{L_{\phi}^2}{\xi_1}
\end{equation}

\noindent and then steeply decreases to zero, so that $\beta
(x)\simeq \beta$ for $x>L_0$. Similarly the coherent term in
eq.~(\ref{rhoL}) reaches maximum at $L=L_0$ and then steeply
decreases to zero. Therefore, $\bar \rho(L)\simeq \beta L-c$ for
$L>L_0$, in accord with Fig.~\ref{Fig5}, where $L_0$ is labeled by
an arrow.

In other words, after leaving the source the Fermi electron first
overcomes the distance $L_0$ at which it acquires the mean
resistance $\bar \rho (L_0)\ll \beta L_0$. Beyond this distance it
contributes ohmically but the shift $c \approx \beta L_0$ is
fixed.

Figure~\ref{Fig5} suggests that $L_{\phi}$ and $\xi_1$ might in
principle be measurable as follows. Measuring the mean resistance
in the linear regime (in long wires) one could determine $\beta$.
One could then extrapolate this linear dependence up to the
intersection point with the $L$-axis, determine $L_0$ as the
$L$-value at the intersection point, and obtain $L_{\phi}$ and
$\xi_1$ from eqs.~(\ref{Blenc}) and~(\ref{L0}). This is justified
if $L_{\phi}/\xi_1 \gg 1$, i.e., if the intersection point and
$L_0$ coincide (in Fig.~\ref{Fig5} these points are still slightly
separated for $L_{\phi} = 10.4$~$\mu$m). One could also directly
measure $\xi_1$ (by detecting the $0.5 \left[ \exp(2L/\xi_1)-1
\right]$ dependence in wires shorter than $L_{\phi}$) and obtain
$L_{\phi}$ from eq.~(\ref{Blenc}). All this is however difficult
in practice, as we discuss below.

\subsection{Is the mean resistance measurable in practice?}

\begin{figure}[t]
\noindent
\centerline{\includegraphics[clip,width=0.85\columnwidth]
{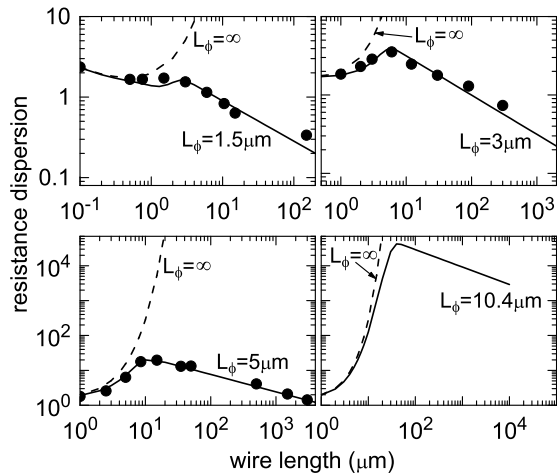}}
 \caption{ Resistance dispersion versus the wire
length $L$ for the same parameters as Fig.~\ref{Fig4}. The dashed
line shows the coherent result~(\ref{rhodispcoh}), dispersion with
decoherence are the full circles (decoherence model of Sect.~IV)
and full lines [eq.~(\ref{rhodisp})]. } \label{FigDisp}
\end{figure}

To address this question we need to calculate the resistance
dispersion $d(L) \equiv (\bar{\rho^2}(L) -
\bar{\rho}^2(L))^{1/2}/\bar{\rho}(L)$. In the coherent regime we
insert for $\bar{\rho}(L)$ and $\bar{\rho^2}(L)$ the
formulae~(\ref{e18}) and~(\ref{e19}), and we obtain the coherent
dispersion

\noindent
\begin{equation} \label{rhodispcoh}
d_{coh}(L) = \sqrt{\frac{1}{3} \left[A-1\right]^{-2}
\left[2B-6A+4\right]-1} ,
\end{equation}

\noindent where $A=\exp \left( 2L/\xi_1 \right)$ and $B=\exp
\left( 6L/\xi_2 \right)$. In Fig.~\ref{FigDisp} the coherent
dispersion~(\ref{rhodispcoh}) is shown in a dashed line. It
increases for large $L$ even faster than its weak-disorder limit
$\exp(L/\xi)$ since our disorder is not weak enough (note in the
caption of Fig.~\ref{Fig4} that $\xi_2<\xi_1<\xi$).

However, this fast increase diminishes in presence of decoherence.
The full circles in Fig.~\ref{FigDisp} show the dispersion
obtained from our decoherence model. It coincides with the
coherent dispersion when $L<L_{\phi}$ , but as $L$ exceeds
$L_{\phi}$ it tends to decrease like $L^{-1/2}$ rather than to
increase. This is easy to understand. Consider an ``incoherent
wire'' consisting of $n$ independent coherent segments of length
$s$. The resistance dispersion of each segment is given by
eq.~(\ref{rhodispcoh}) as $d_{coh}(s)$ and the resistance
dispersion of such wire is simply $d_{incoh}=n^{-1/2}d_{coh}(s) =
(s/L)^{1/2} d_{coh}(s)$, i.e., $d_{incoh}(L) \propto L^{-1/2}$.

It seems reasonable to improve this simple model as

\noindent
\begin{equation} \label{rhodispincoh}
d_{incoh}(L) = \sqrt{\frac{\bar s}{L}} \int \limits _0 ^L ds \, g
\left(s\right) d_{coh}\left(s\right) ,
\end{equation}

\noindent where $g \left(s\right)$ is given by eq.~(\ref{PDelta})
and $\bar s=\int_{0}^{\infty} ds \, g \left(s\right) s$, and to
express the resistance dispersion of the wire as

\noindent
\begin{equation} \label{rhodisp}
d(L) = e^{- \frac{L^2}{L_{\phi}^2}} \ d_{coh} \left(L\right) + (
1- e^{- \frac{L^2}{L_{\phi}^2}} ) \ d_{incoh}(L) ,
\end{equation}

\noindent where $\exp{(-L^2/L_{\phi}^2)}$ is the coherent
transmission probability. In Fig.~\ref{FigDisp} the
dependence~(\ref{rhodisp}) is shown in the full line, its
agreement with our decoherence model results is good. The
decoherence model needs a lot of computational time to give the
resistance dispersion with a reasonable accuracy and this time
increases very fast with increasing $L_{\phi}/\xi_1$. Therefore,
in Fig.~\ref{FigDisp} no full circles are shown for  $L_{\phi} =
10.4$~$\mu$m, but we believe that eq.~(\ref{rhodisp}) gives a
reasonable information also in this case.

To measure $\bar{\rho}(L)$ in practice, one needs to measure the
resistance of the ensemble of wires, of course, each wire in the
ensemble has to be biased separately. Ideally, the wires should be
identical as to their geometry, doping level, electron density,
etc., and the number of the occupied 1D channels should be tunable
up to the ground energy subband. Finally, the distance between the
nearest wires should be say a few microns to exclude the
electron-electron inter-wire coupling. All together, these
conditions are difficult to realize in practice, but experiments
with the ensembles of several tens to several hundreds of such
wires might perhaps be realizable.

How many wires are needed to obtain a reliable ensemble average?
Assume \cite{error} that $\bar{\rho}(L)$ is measured with the
error $d(L)/\sqrt{n_w}$, where $n_w$ is the number of wires in the
ensemble. To achieve for instance $d(L)/\sqrt{n_w} \approx 1/2$,
in case of Fig.~\ref{FigDisp} we need $n_w \simeq 1-500$ for
$L_{\phi}=1.5-5$~$\mu$m while for $L_{\phi} = 10.4$~$\mu$m the
upper limit of $n_w$ as large as $10^{10}$ is needed. Generally
speaking, $n_w \simeq 1-500$ suffices if $L_{\phi}/\xi_1<2$, for
larger $L_{\phi}/\xi_1$ the upper limit of $n_w$ is simply too
large to be realizable in practice. In the latter case
$\bar{\rho}(L)$ is still measurable say for $d(L)<10$, i.e., in
the very short or very long wires.

To measure the resistance of millions of disordered wires it would
be desirable to fabricate a single quantum wire with tunable
disorder. We have in mind the wire in which intrinsic disorder is
negligible but a tunable extrinsic disorder is introduced by using
an array of very narrow equidistant gates biased separately by a
voltage varying at random from gate to gate. Biasing a single gate
creates a localized potential barrier inside the wire underneath
the gate. Thus, the array of the gates of width $\sim 2 \pi/k_F$
would create disorder of equidistant random barriers, each of them
with the quantum-reflection coefficient $R_I$ tunable between $0$
and $1$ by external bias.

\begin{figure}[t]
\noindent
\centerline{\includegraphics[clip,width=0.85\columnwidth]
{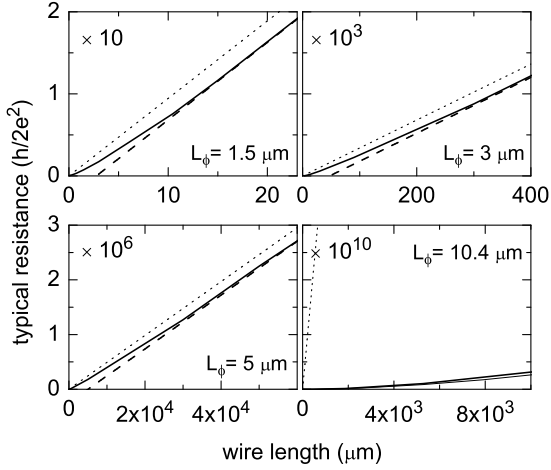}}
\caption{The typical resistance from Fig.~\ref{Fig4}
shown in a linear scale: the dotted line shows the dependence
$\beta L$, the full line is the typical resistance resulting from
the decoherence model of Sect.~IV, the dashed line is the shifted
$\beta L$ dependence to which the typical resistance converges.
The resistance data should be multiplied by a factor $1$, $10^3$,
$10^6$, or $10^{10}$, as indicated. The thin full line is the
approximation of the typical resistance described by
eqs.~(\ref{rhoT}-\ref{fseriaGauss}). } \label{Fig9}
\end{figure}

Perhaps a proper candidate for realization of tunable disorder is
the cleaved-edge overgrowth quantum wire system, which shows
ballistic transport (negligible intrinsic disorder) for wires as
long as several tens of microns. \cite{Yacoby-96} One could
speculate about an array of gates, or instead of the gate array
one could think about an array of separately biased non-invasive
probes. Such probes were already realized in the cleaved-edge
overgrowth wire \cite{dePicciotto-01} in a different context, they
allow to tune also the wire length.

\subsection{Typical resistance and distribution of $\ln(1+\rho)$}

The typical resistance is defined as $\rho_t=\exp(\bar f)-1$,
where $\bar f$ is the ensemble average of the variable
$f=\ln(1+\rho)$. Obviously, $\rho_t=\bar \rho$ if $\rho=\bar
\rho$, i.e., if there are no resistance fluctuations. Since the
resistance fluctuations decrease in presence of decoherence as
$1/\sqrt{L}$ (c.f. Fig.~\ref{FigDisp}), one might expect the
typical resistance to approach the mean resistance with increasing
$L$. Reality is however more complicated.  We have seen in
Fig.~\ref{Fig4} that decoherence drives the typical resistance
towards the same linear slope as it drives the mean resistance. In
Fig.~\ref{Fig9} we present the typical resistance again, but in a
linear scale. Clearly, decoherence drives the typical resistance
towards the linear dependence $\beta L-c^\prime$, where the shift
$c^\prime$ strongly differs from the shift $c$ of the mean
resistance (c.f. Fig.~\ref{Fig5}). The lines $\rho_t(L) \propto L$
and $\bar \rho(L) \propto L$ remain parallel, i.e., $\rho_t$
differs from $\bar \rho$ also in the ohmic regime.

To give insight we wish to assess $\rho_{t}(L)$ more
transparently. In analogy with eqs.~(\ref{rhoL})
and~(\ref{betax}), we approximate

\noindent
\begin{equation} \label{rhoT}
\rho_{t}(L) \approx
 \rho(L)  \,
e^{- \frac{L^2}{L_{\phi}^2}} + (1-e^{- \frac{L^2}{L_{\phi}^2}})
 \int \limits _0 ^L dx \beta_{t}(x) ,
\end{equation}

\noindent
\begin{equation} \label{betaxT}
\beta_{t} (x) =  \frac{ \int \limits _0 ^x ds \, g \left(s \right)
\rho(s) } {
  \int \limits _0 ^x ds \, s \, g \left(s \right)
} ,
\end{equation}

\noindent
but we assess $\rho(s)$ as a typical resistance of the
segment.

Imagine a segment $s$ positioned in the beginning of a wire with
specific disorder. If we move the segment along the wire,
$\rho(s)$ fluctuates discretely each time an impurity leaves or
enters the segment. However, since a single impurity scatters
weakly, differences between the successive discrete values can be
neglected except for resonant changes occurring (on average)
whenever the segment is shifted by $\xi$. The number of such
changes,

\noindent
\begin{equation} \label{nsegments}
n = \text{Int} \left [\frac{x-s}{\xi} \right] + 1 ,
\end{equation}

\noindent results in a serie of $n$ different values of $\rho(s)$.
Denoting this serie as $\rho_{1}(s)$, $\rho_{2}(s)$, $\dots$,
$\rho_{n}(s)$ we can write

\noindent
\begin{equation} \label{rhoS}
\rho (s) = \frac{1}{n}\, \sum_{i=1}^n \rho_{i}(s) = \frac{1}{n}\,
\sum_{i=1}^n ( e^{f_{i}}-1 ) ,
\end{equation}

\noindent where the variable $f_{i} = \ln (1 + \rho_{i}(s))$
follows the coherent distribution ${\cal P}(f,s)$ discussed in
Sects.~II--III. Each $f$ in a specific serie $f_{1}(s)$,
$f_{2}(s)$, $\dots$ $f_{n}(s)$ can thus be viewed as a random
number from that distribution. The so-called typical serie
$f_{1}$, $f_{2}$, $\dots$, $f_{n}$ is given by equations

\noindent
\begin{equation} \label{fseria}
\frac{1}{n+1} = \int \limits _0 ^{f_{1}} df {\cal P}(f) = \int
\limits _{f_{1}}^{f_{2}} df {\cal P}(f) = \dots =\int \limits
_{f_{n-1}}^{f_{n}} df {\cal P}(f),
\end{equation}

\noindent because one expects that the total area $\int _{0}
^{\infty} df {\cal P}(f)=1$ is on average divided into $n+1$ areas
of size $1/(n+1)$ (the argument $s$ is skipped for simplicity).
Using

\noindent
\begin{equation} \label{distfs}
{\cal P}(f) = \frac{1}{\sqrt{2\pi\Delta^2}}
\exp\left[-\frac{{(f-\bar{f})}^2}{2\Delta^2}\right], \qquad \bar f
= s/\xi
\end{equation}

\noindent we get from eq.~(\ref{fseria}) the equations

\noindent
\begin{equation} \label{fseriaGauss}
\text{erf} \left(\frac{f_i - \bar f}{\sqrt{2\Delta^2}}\right) =
\frac{2i}{n+1} -1 , \qquad i = 1, \dots, n .
\end{equation}

\noindent Equation~(\ref{distfs}) is an approximation valid for
$s/\xi \gg 1$. It roughly holds for any $s/\xi$, if one adjusts
\cite{Anderson-80} $\Delta^2$ to fulfill the equation $\int df
{\cal P}(f) (\exp(f)-1) = 0.5 \left[ \exp(2s/\xi_1)-1 \right]$.
This gives $\Delta^2 = 2s/\xi_1 - 2s/\xi + 2 \ln \cosh(s/\xi_1)$.

It is easy to see that eqs.~(\ref{rhoT}-\ref{fseriaGauss}) give
for large $L$ $\rho_{t}(L) = \beta L-c^\prime $, where $c^\prime >
c$. The segment resistance~(\ref{rhoS}) increases with $n$, its
lower limit at $n=1$ is $ \exp(f_1)-1 = \exp(L/\xi)-1$ while its
upper limit at large $n$ is $\int df {\cal P}(f)(\exp(f)-1) = 0.5
\left[ \exp(2s/\xi_1)-1 \right]$. Physically, eq.~(\ref{rhoS}) is
the self-average of $\rho(s)$ in a single ``typical'' wire and
therefore coincides with the ensemble average $0.5 \left[
\exp(2s/\xi_1)-1 \right]$ only at very large $x$, at which it 
involves also the tail of the resistance distribution. Otherwise
$\rho(s) \ll 0.5 \left[ \exp(2s/\xi_1)-1 \right]$ and consequently
$\beta_{t}(x) \ll \beta (x)$. Eventually, $\beta_{t}(x)$ converges
towards $\beta $, but at much larger $x$ as does the $\beta (x)$,
so that $\rho_{t}(L)=\beta L-c^\prime $ with the shift $c^\prime
\gg c$. In other words, the shift $c^\prime $ reflects a slowly
converging process of the resistance self-averaging in a single
wire.

We have also extracted $\rho_{t}(L)$ from
eqs.~(\ref{rhoT}-\ref{fseriaGauss}) numerically. The result obtained
for $L_{\phi} = 10.4$~$\mu$m is shown in Fig.~\ref{Fig9} in a thin
full line. It reasonably fits the decoherence model result shown in
the thick full line, the fit further improves with increasing
$L_{\phi}/\xi_1$ (not shown). Note that for $L_{\phi} = 10.4$~$\mu$m
the slope of the dotted line strongly differs from the slope of the
thick full line, i.e., here the thick full line is quite far from the
ohmic regime. It was not traced up to the ohmic regime, since the
decoherence model of Sect.~IV needs in this case too much
computational time.

We have therefore applied the
scheme~(\ref{rhoT}-\ref{fseriaGauss}), which is computationally
fast. We have found that for $L_{\phi} = 10.4$~$\mu$m the typical
resistance tends to be linear at $L>10^{11}$~$\mu$m. We cannot
offer any simple formula to explain this unprecedently large
value, but it is interesting to compare Figs.~\ref{FigDisp}
and~\ref{Fig9}. It can be seen that for $L_{\phi} = 1.5$, $3$, and
$5$~$\mu$m the typical resistance becomes linear at those $L$ at
which the resistance dispersion becomes smaller than unity.
Similarly, the dispersion for $L_{\phi} = 10.4$~$\mu$m would reach
unity nearly at $L=10^{11}$~$\mu$m.

\begin{figure}[t]
\noindent
\centerline{\includegraphics[clip,width=0.85\columnwidth]
{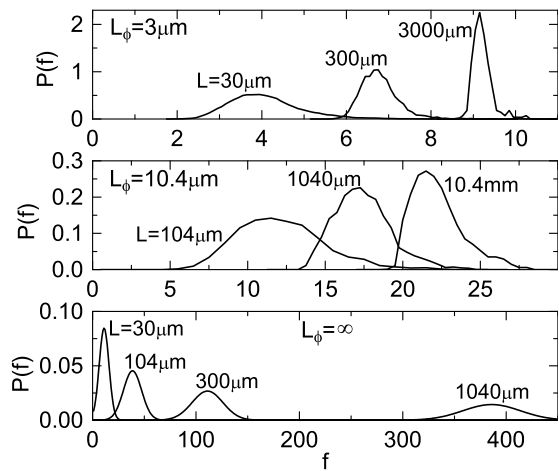}} \caption{ Distribution of the variable $f = \ln (1 +
\rho )$, ${\cal P}(f)$, for various wire lengths $L$.
Distributions in the coherent regime ($L_{\phi}=\infty$) are
compared with the distributions perturbed by decoherence
($L_{\phi}=3$ $\mu$m and $L_{\phi}=10.4$ $\mu$m). } \label{Fig6}
\end{figure}

The typical resistance should be much easier to measure than the
mean resistance, because the dispersion of the variable $f$ is
small already in the coherent regime ($\Delta/\bar{f} \approx
1/\sqrt{L/\xi}$ for $L/\xi \gg 1$). In presence of decoherence the
situation is favourable as well. Figure~\ref{Fig6} shows the
distributions ${\cal P}(f)$ obtained in the same simulation as the
results of Fig.~\ref{Fig4}. It can be seen that the coherent
distribution broadens with increasing $L$ while the decoherence
narrows the distribution as $L$ exceeds $L_{\phi}$. Eventually, at
$L\gg L_{\phi}$ the decoherence drives the distribution towards a
$\delta$-function-like shape, i.e., the dispersion of $f$ tends to
be suppressed.

\section{Summary and discussion}

We have modelled the effect of phase-breaking electron collisions
on the electron transport in a disordered 1D wire with single
conducting channel. In our model the phase-breaking collisions
effectively break the wire into independent segments, where each
segment is a series resistor with coherent electron motion. We
have modeled the segmentation as a stochastic process with a
Poisson distribution of phase-breaking scattering times and we
have evaluated the resistance of each segment microscopically from
the Landauer formula.

The wire resistance as a function of the wire length $L$,
coherence length $L_{\phi}$, and localisation length $\xi$ has
been traced from the coherent regime at $L\!<\!L_{\phi}$ up to the
incoherent one at $L\!\gg\!L_{\phi}$.

In the coherent regime the resistance fluctuates from wire to wire
in accord with the DMPK resistance distribution, the typical
resistance increases as $\exp(L/\xi)$, and the mean resistance
increases as $\exp(2L/\xi)$. This is in agreement with the
universal scaling theory.

As $L$ exceeds $L_{\phi}$, decoherence suppresses the resistance
fluctuations and narrows the resistance distribution. As a result,
at $L\!\gg\!L_{\phi}$ the mean resistance increases as $\beta L-c$
and the typical resistance as $\beta L- c^\prime $, where $\beta$
is the wire resistivity, $c$ is a constant shift due to the
decoherence near the source contact, and $c^\prime \gg c$ is the
shift related to the resistance self-averaging in a single wire.

As argued in Sect.~IV, our model is in essence a simple
phenomenological model of the e-e interaction mediated phase
breaking, with the electron coherence length $L_{\phi}$ taken as a
parameter. If compared with experiment, the model could help to
determine the coherence length and also the localization length.

We have also discussed conditions necessary to measure our results
in practice. The mean resistance might be measurable if $L_{\phi}$
does not exceed $\xi $ more than about twice, otherwise the
resistance fluctuations become too large to perform average over a
sufficient number of wires. To average over millions of wires, one
could perhaps speculate about a single wire with tunable disorder
induced by an array of randomly biased gates. As to the typical
resistance, one fortunately needs to average the variable $f = \ln
(1 + \rho)$, which is much easier to do.

We add a comment about the effect of finite temperature. We have
implicitly assumed that the temperature is low enough for the
contribution from the phonon mediated dephasing to be negligible.
In the GaAs systems this is justified below $\sim 1$~K. However,
we have used the zero-temperature formula~(\ref{e1}), which may
not be appropriate even for $T \ll \varepsilon_F /k_B$. We have
therefore tested for $T \ll \varepsilon_F /k_B$, what happens if
we replace the formula~(\ref{e1}) by its temperature-dependent
version. \cite{Buttiker-99} We have found that in case of coherent
transport the resistance distribution and the mean and typical
resistance depend on the localisation length as at zero
temperature, except that the localisation length is $T$-dependent.
\cite{Kravtsov} The same has to hold for each coherent segment in
a wire segmented by decoherence, thus the results presented in
this paper are representative also for non-zero temperatures.

Our numerical simulations were performed in conditions
characteristic of the GaAs quantum wire. In such wire the
backscattering by disorder can be strong. \cite{Mosko-99} We have
therefore also examined the effect of strong disorder on coherent
transport. We have found deviations from universal scaling which
differ from those reported for the Anderson model.
\cite{Slevin-90}

\acknowledgments P. V. was supported by a Marie Curie Fellowship
of the Fifth Framework Programme of the European Community,
contract no. HPMFCT-2000-00702. M.M. and P.V. were also supported
by the VEGA grant no. 2/7201/21 and P. M. by the VEGA grant no.
2/7174/20.

\appendix
\section{}

The purpose of this Appendix is to analyze the coherent resistance
of the chain of $N$ identical obstacles and to derive the
formulae~(\ref{e15}),~(\ref{e16}), and~(\ref{e17}).

\subsection{Mean resistance}

The mean resistance~(\ref{e15}) was derived by Landauer,
\cite{Landauer-70} but it can be interesting to derive the same
result in a more simple way. Consider two obstacles with
resistances $\rho_a=R_a/(1-R_a)$ and $\rho_b=R_b/(1-R_b)$, where
$R_a$ and $R_b$ are the reflection probabilities of the obstacles.
The composite resistance of the two obstacles is
\cite{Landauer-70}

\noindent
\begin{equation} \label{AppRho12}
\rho_2 = \frac{R_a + R_b + 2 \sqrt{R_a R_b} \cos \phi} {(1-R_a)
(1-R_b)} ,
\end{equation}

\noindent where the phase $\phi = 2k_{F}a + \phi_0$, $a$ is the
interobstacle distance, and $\phi_0$  is the ($a$-independent)
phase shift due to the reflection by the obstacles . If one
assumes that $a \gg 2\pi/k_F$, then $\phi$ changes rapidly with
$a$ and fluctuates at random from sample to sample as $a$
fluctuates. In such case the ensemble average of
eq.~(\ref{AppRho12}) over the interobstacle distance is simply

\noindent
\begin{equation} \label{Apprho}
\bar \rho_{2} = \frac{1}{2\pi} \, \int \limits_0 ^{2\pi} d\phi \,
\rho_2 .
\end{equation}

\noindent Inserting (\ref{AppRho12}) into (\ref{Apprho}) one finds

\noindent
\begin{equation} \label{Apprho2}
\bar\rho_{2} = \frac{R_a} {1-R_a} + \frac{1+R_a} {1-R_a} \,
\frac{R_b} {1-R_b} .
\end{equation}

\noindent Consider now a chain of $N$ identical obstacles, say
impurities, where each obstacle is characterized by the reflection
probability $R_I$. We follow the scaling procedure of
Ref.~\onlinecite{Landauer-70}: We identify $R_a$ with the
reflection probability of a single impurity added to the chain of
$N-1$ impurities, we average over all phase differences in the
chain of these $N-1$ impurities, and we identify $R_b/(1-R_b)$
with the mean resistance $\bar \rho_{N-1}$.
Equation~(\ref{Apprho2}) then gives

\noindent
\begin{equation} \label{AppwN}
\bar\rho_N = \frac{\alpha-1}{2} + \alpha\, \bar \rho_{N-1} ,
\end{equation}

\noindent where $\alpha=(1+R_I)/(1-R_I)$. We rewrite
eq.~(\ref{AppwN}) as

\noindent
\begin{equation} \label{AppRecur}
\bar\rho_{N} = \frac{\alpha-1}{2} \, \left( 1+ \alpha + \alpha^2 +
\dots + \alpha^{N-1} \right) ,
\end{equation}

\noindent which immediately gives eq. (\ref{e15}) as

\noindent
\begin{equation} \label{AppLandFinal}
\bar{\rho}(N)= \frac{\alpha-1}{2} \, \frac{\alpha^N-1}{\alpha-1} =
\frac{1} {2} \left[ {\left(\frac{1+R_I}{1-R_I}\right)}^N-1 \right]
.
\end{equation}

\subsection{Typical resistance}

To obtain the typical resistance~(\ref{e17}), one has to average
over all possible configurations of impurity positions the
variable $f_N= \ln \left( 1+\rho_N \right)$, where $N$ is the
number of impurities. We start with two impurities and perform the
ensemble average of $f_2=\ln(1+\rho_2)$, where $\rho_2$ is given
by the formula~(\ref{AppRho12}). Similarly as in the preceding
subsection, the ensemble average of $f_2$ can be written as

\noindent
\begin{equation} \label{AppLandTyp1}
\bar f_{2} = \frac{1}{2\pi} \, \int \limits_0 ^{2\pi} d\phi \, \ln
\left[ \frac{1 + R_a R_b + 2 \sqrt{R_a R_b} \cos \phi} {(1-R_a)
(1-R_b)} \right] ,
\end{equation}

\noindent which gives

\noindent
\begin{equation} \label{AppLandTyp2}
\bar f_2 = \ln \left( \frac{1}{1-R_{a}} \right) + \ln \left(
\frac{1}{1-R_{b}} \right) .
\end{equation}

\noindent Applying the scaling procedure of the preceding
subsection one easily obtains the recursion relation

\noindent
\begin{equation} \label{AppLandTyp3}
\bar f_N = \ln \left( \frac{1}{1-R_I} \right) + \bar f_{N-1} ,
\end{equation}

\noindent which gives

\noindent
\begin{equation} \label{AppLandTyp4}
\bar f_N = N \ln \left( \frac{1}{1-R_I} \right) .
\end{equation}

\noindent Finally, inserting eq.~(\ref{AppLandTyp4}) into the
definition $\rho_t \left( N \right) = e^{\bar f_N}-1 $ we obtain
the typical resistance~(\ref{e17}).

\subsection{Mean squared resistance}

Now we derive eq.~(\ref{e16}), i.e., we calculate the mean value
$\bar{\rho^2}(N)$ of the squared resistance $\rho^{2}(N)$. We
start again with two impurities and calculate the quantities

\begin{subequations} \label{AppAllQuant}
\noindent
\begin{equation} \label{AppX2Def}
\bar X_{2} = \frac{1}{2\pi} \, \int \limits_0 ^{2\pi} d\phi \,
\left( \rho_2 \right)^2 ,
\end{equation}

\noindent
\begin{equation} \label{AppY2Def}
\bar Y_{2} = \frac{1}{2\pi} \, \int \limits_0 ^{2\pi} d\phi \,
\left( 1+\rho_2 \right)^2 ,
\end{equation}

\noindent and

\noindent
\begin{equation} \label{AppZ2Def}
\bar Z_{2} = \frac{1}{2\pi} \, \int \limits_0 ^{2\pi} d\phi \, 2
\rho_2 \left( 1+\rho_2 \right) .
\end{equation}
\end{subequations}

\noindent where $\rho_2$ is given by (\ref{AppRho12}). We obtain a
set of equations

\noindent
\begin{subequations} \label{AppAllEq}
\begin{equation} \label{AppAllEqX}
\bar X_2 = Y_a X_b + X_a Y_b + Z_a Z_b ,
\end{equation}

\noindent
\begin{equation} \label{AppAllEqY}
\bar Y_2 = X_a X_b + Y_a Y_b + Z_a Z_b ,
\end{equation}

\noindent
\begin{equation} \label{AppAllEqZ}
\bar Z_2 = Z_a X_b + Z_a Y_b + \left( X_a+Y_a+Z_a \right) Z_b ,
\end{equation}
\end{subequations}

\noindent where $X_a=[R_a/(1-R_a)]^2$, $Y_a = 1/(1-R_a)^2$, $Z_a
=2R_a/(1-R_a)^2$, and $X_b$, $Y_b$, and $Z_b$ are given
analogously.

We apply the scaling procedure of subsection~1 and rewrite
eqs.~(\ref{AppAllEq}) into the recursion vector form

\noindent
\begin{equation} \label{AppEqVect}
\left(
  \begin{array}{c}
    \bar X_N \\ \bar Y_N \\ \bar Z_N
  \end{array}
\right) = {\mathcal{A}} \, \left(
  \begin{array}{c}
    \bar X_{N-1} \\ \bar Y_{N-1} \\ \bar Z_{N-1}
  \end{array}
\right) = {\mathcal{A}}^{N-1} \, \left(
  \begin{array}{c}
    X_I \\ Y_I \\ Z_I
  \end{array}
\right) ,
\end{equation}

\noindent where

\noindent
\begin{equation} \label{AppEqA}
{\mathcal{A}}= \left(
  \begin{array}{ccc}
    Y_I & X_I & Z_I \\
    X_I & Y_I & Z_I \\
    Z_I & Z_I & X_I+Y_I+Z_I \\
  \end{array}
\right) .
\end{equation}

\noindent Equations~(\ref{AppEqVect}) can be solved by expanding
the vector $(X_I,Y_I,Z_I)$ as a linear combination of eigenvectors
of the matrix $\mathcal{A}$. We obtain

\noindent
\begin{equation} \label{AppFinalX1}
\bar X_N = -\frac{1}{2}\, \lambda_1^N +\frac{1}{3}\, \lambda_2^N
+\frac{1}{6}\, \lambda_3^N ,
\end{equation}

\noindent
where $\lambda_1$, $\lambda_2$, and $\lambda_3$ are the
eigenvalues of $\mathcal{A}$, namely $\lambda_1 = Y_I-X_I$,
$\lambda_2 = X_I+Y_I-Z_I$, and $\lambda_3 = X_I+Y_I+2Z_I$. In
terms of $R_I$ we have $\lambda_1 =(1+R_I)/(1-R_I)$, $\lambda_2 =
1$, and $\lambda_3 =1 + 6R_I/( 1-R_I)^2$. Inserting these
expressions into eq.~(\ref{AppFinalX1}) we obtain the equation

\noindent
\begin{equation} \label{AppFinalX}
\bar{\rho^2}(N) = \frac{1}{3} -\frac{1}{2} \left( \frac{1+R_I}
{1-R_I} \right)^N +\frac{1}{6} \left( 1 + \frac{6R_I} {\left(
1-R_I \right)^2} \right)^N ,
\end{equation}

\noindent
which is identical with eq.~(\ref{e16}).



\begin{thebibliography}{99}
\bibitem{commentPdelta}
For example, occurence of the segment $s = 0.7L$ in the wire of
length $L$ implies that occurence of the segments $s > L-0.7L$ in
the same configuration of segments is not allowed.
Eqs.~(\ref{betax}) and~(\ref{PDelta}) ignore this problem.
\bibitem{error}
The formula $d(L)/\sqrt{n_w}$ is applicable if $d(L) \propto
1/\sqrt{L}$, which is the case in presence of decoherence. In the
coherent regime it is not applicable if $L \gg \xi$ (according to
our numerical experience  if $d_{coh}(L)>10$), because for $L \gg
\xi$ the $d_{coh}(L)$ is completely dominated by the tail of the
resistance distribution.
\end{thebibliography}
\end{document}